# THERMODYNAMICS OF 2-UNDECANONE + *N*-ALKANE MIXTURES

Susana Villa, Fatemeh Pazoki, Juan Antonio González*, Fernando Hevia, Daniel Lozano-Martín, Luis Felipe Sanz, João Victor Alves-Laurentino

G.E.T.E.F., Departamento de Física Aplicada, Facultad de Ciencias, Universidad de Valladolid, Paseo de Belén, 7, 47011 Valladolid, Spain.

*corresponding author, e-mail: jagl@termo.uva.es

**Abstract**

Densities and excess molar volumes ($V_\mathrm{m}^\mathrm{E}$) at (293.15-303.15) K and excess molar enthalpies ($H_\mathrm{m}^\mathrm{E}$) at 298.15 K are reported for 2-undecanone + heptane, or + octane, or + decane, or + dodecane, or + tetradecane mixtures at 95 kPa. Densities and $H_\mathrm{m}^\mathrm{E}$ were measured using, respectively, a densitometer Anton Paar DMA 602 and a Tian-Calvet micro-calorimeter. $H_\mathrm{m}^\mathrm{E}$ results are positive, indicating that interactions between like molecules are dominant. Both $H_\mathrm{m}^\mathrm{E}$ and $V_\mathrm{m}^\mathrm{E}$ increase in line with $n$ (number of C atoms of the alkane), which reveals that the increase of $V_\mathrm{m}^\mathrm{E}$ can be ascribed to an increased interactional contribution. Nevertheless, systems with $n$ = 7,8 show negative $V_\mathrm{m}^\mathrm{E}$ values, which reveals that $V_\mathrm{m}^\mathrm{E}$ is determined mainly by structural effects. Isochoric excess molar internal energies ($U_{V\mathrm{m}}^\mathrm{E}$) at 298.15 K have been obtained from the present $H_\mathrm{m}^\mathrm{E}$ and $V_\mathrm{m}^\mathrm{E}$ data. At equimolar composition, $U_{V\mathrm{m}}^\mathrm{E}$ is nearly constant for $n$ = 7-10, and then slightly increases. This has been explained in terms of a possible folding of 2-undecanone. From the comparison of $U_{V\mathrm{m}}^\mathrm{E}$ results for similar systems involving $n$-alkanoates, it is shown that folding is more likely in solutions with these compounds.

**Keywords:** 2-undecanone, alkanes, excess molar functions, folding

## 1. Introduction

2-Alkanones are polar, aprotic compounds and hydrogen bond acceptors with many different applications. They are used as solvents for plastics and for some synthetic fibers or as intermediates for obtaining very important compounds, such as methyl methacrylate. They have also an essential role in biochemistry since many sugars are ketones, and the synthesis of fatty acids is carried out using these compounds. On the other hand, the research of (alkanone + alkane) mixtures allows the study of the impact of a number of effects on the thermodynamic properties of the considered systems. Thus, it is possible to investigate among others: (i) the effect of increasing the alkyl chain of *n*-alkanones while the position of the CO group remains unchanged within the alkyl chain; (ii) isomeric effects related to the different position of the CO group within a given alkyl chain; (iii) cyclization or aromatic effects by means of the study of solutions with cycloalkanones or aromatic ketones, respectively. In this framework, we have investigated previously orientational effects in (alkanone + alkane) mixtures and the aromaticity effect [1-4] from the application of different models, such as Flory [5] or DISQUAC [6], and the measurement of liquid-liquid equilibria. Another effect which can be also investigated is the possible folding of *n*-alkanones in their systems with *n*-alkanes. The possibility that *n*-alkanones or *n*-alkanoates can adopt quasi-cyclic structures in *n*-alkane solutions was suggested earlier by Piekarski et al. on the basis of spectroscopic and thermophysical data [7-10]. In a previous work [11], we have confirmed that *n*-alkanoates can adopt folded structures from the determination of isochoric excess internal energies ($U_{V\mathrm{m}}^{\mathrm{E}}$) and from volumetric data. Now, we extend that study to *n*-alkanone mixtures and we provide calorimetric and volumetric data for the systems 2-undecanone + heptane, or + octane, or + decane, or + dodecane, or + tetradecane. Particularly, we have measured excess molar enthalpies ($H_{\mathrm{m}}^{\mathrm{E}}$) at 298.15 K and densities ($\rho$) over the temperature range (293.15-303.15) K. From the latter values, excess molar volumes ($V_{\mathrm{m}}^{\mathrm{E}}$) were determined. All the measurements were performed at atmospheric pressure (95 kPa). This set of data has been used for obtaining $U_{V\mathrm{m}}^{\mathrm{E}}$ values at 298.15 K.

## 2. Experimental

### *2.1. Materials*

Pure liquids were used as received from the supplier, without any further purification. Table 1 contains the source and purity of the chemicals. Solutions were prepared by weighing in small vessels of about 10 cm$^3$. The concentration of the mixtures (given by the mole fraction of 2-undecanone, $x_1$) was calculated from mass measurements carried out using an analytical balance (MSU125P, Sartorius) and correcting for buoyancy effects. The standard uncertainty of these measurements is $5\cdot10^{-5}$ g. Along the process, caution was taken in order to prevent evaporation.

Conversion to molar quantities was based on the relative atomic mass Table of 2015 issued by I.U.P.A.C [12]. The standard uncertainty in the final mole fraction is estimated to be 0.0005. All measurements were performed at 95 kPa. In addition, the temperature (standard uncertainty of 0.01 K) of the samples was measured by means of Pt-100 resistances, calibrated using the triple point of water and the melting point of Ga.

**TABLE 1**

Sample description.

| Chemical name | CAS Number | Source | Initial purity[a] | Purification method |
|---|---|---|---|---|
| 2-undecanone | 112-12-9 | Sigma-Aldrich | 0.994 | none |
| heptane ($n$-$C_7$) | 142-82-5 | Sigma-Aldrich | 0.998 | none |
| octane ($n$-$C_8$) | 111-65-9 | Sigma-Aldrich | 0.997 | none |
| decane ($n$-$C_{10}$) | 124-18-5 | Sigma-Aldrich | 0.995 | none |
| dodecane ($n$-$C_{12}$) | 112-40-3 | Sigma-Aldrich | 0.998 | none |
| tetradecane ($n$-$C_{14}$) | 629-59-4 | Fluka | 0.995 | none |

[a]Gas chromatography area fraction, certified by the supplier.

### 2.2. *Density measurements*

Densities of the liquid samples were measured by means of a densitometer Anton Paar DMA 602, whose temperature stability was 0.01 K. For density measurements, the standard uncertainty under repeatability conditions is estimated to be $5\times10^{-2}$ kg m$^{-3}$, which gives criteria for the number of significant figures. As usual, several measurements of the densities ($\rho_i$) of the pure liquids were conducted over time in order to test their stability, remaining constant within the measurement uncertainty. The apparatus was calibrated using the following reference pure liquids: heptane, isooctane, cyclohexane, toluene and water HPLC plus. Table S1 (supplementary material) contains the source and purity of these compounds. The procedure was applied at each of the three working temperatures. More details on the calibration and on the applied procedure can be encountered elsewhere [13-15]. A comparison between our $\rho_i$ values at (293.15-303.15) K with literature results is shown in Table 2. Differences are lower than 0.1 %, so the total relative combined expanded uncertainty (0.95 level of confidence) can be estimated as 0.002. The densitometer was tested by measuring $V_m^E$ for the reference system (cyclohexane + benzene) at 298.15 K [16]. Our measurements are included in Table S2 (see Figure S1). Although the differences between literature values and our results is lower that 1%, the uncertainty for $V_m^E$ is estimated to be $0.010\left|V_m^E\right|_{max} + 0.005$ cm$^3$ mol$^{-1}$ ($\left|V_m^E\right|_{max}$ denotes the maximum experimental value of $\left|V_m^E\right|$ with respect to the mole fraction).

**TABLE 2**

Densities, $\rho_i$, of pure compounds at temperature $T$ and 95 kPa[a]. Comparison of experimental values (Exp) with results from the literature (Lit).

| Compound | $\rho_i$ /g cm$^{-3}$ | | | | | |
|---|---|---|---|---|---|---|
| | $T$/K = 293.15 | | $T$/K = 298.15 | | $T$/K = 303.15 | |
| | Exp | Lit | Exp | Lit | Exp | Lit |
| 2-undecanone | 0.82610 | 0.8250[44] | 0.82228 | 0.82231[45] | 0.81849 | 0.8184[46] |
| | | 0.8258[46] | | 0.8221[47] | | 0.8183[47] |
| | | | | 0.82170[48] | | |
| | | | | 0.82230[49] | | |
| $n$-C$_7$ | 0.68384 | 0.68382[50] | 0.67959 | 0.67960[50,51] | 0.67533 | 0.67534[50] |
| | | 0.68396[52] | | 0.67951[53] | | 0.67550[52] |
| | | 0.6841[54] | | 0.6795[54] | | 0.6752[54] |
| | | 0.68375[55] | | 0.67952[56] | | 0.67525[55] |
| | | | | 0.67977[52] | | |
| $n$-C$_8$ | 0.70294 | 0.7025[57] | 0.69895 | 0.6985[57] | 0.69488 | 0.6945[57] |
| | | 0.70276[58] | | 0.69873[58] | | 0.69469[58] |
| | | 0.7030[59] | | 0.6989[59,60] | | 0.6949[60] |
| $n$-C$_{10}$ | 0.73006 | 0.73009[50] | 0.72628 | 0.72631[50] | 0.72251 | 0.72253[50] |
| | | 0.7299[61] | | 0.7262[61] | | 0.7225[59] |
| | | 0.72991[52] | | 0.72616[52] | | 0.72236[52] |
| | | 0.72987[62] | | 0.72607[53] | | 0.72231[62] |
| | | 0.73015[63] | | 0.72636[63] | | 0.72247[63] |
| $n$-C$_{12}$ | 0.74896 | 0.74896[50] | 0.74532 | 0.74534[50] | 0.74170 | 0.74172[50] |
| | | 0.7488[61] | | 0.7453[61] | | 0.7417[61] |
| | | 0.74866[64] | | 0.74540[56] | | 0.74138[64] |
| | | 0.74878[52] | | 0.74503[64] | | 0.74152[52] |
| | | 0.74879[62] | | 0.74514[52] | | 0.74154[62] |
| $n$-C$_{14}$ | 0.76285 | 0.76281[50] | 0.75930 | 0.75929[50] | 0.75579 | 0.75576[50] |
| | | 0.7635[61] | | 0.7598[61] | | 0.7563[61] |
| | | 0.76333[65] | | 0.75935[56] | | 0.75630[65] |
| | | 0.76271[62] | | 0.75980[65] | | 0.75565[62] |

[a] The standard uncertainties are: $u(T)$ = 0.01 K; $u(p)$ = 10 kPa. The relative combined expanded uncertainty (0.95 level of confidence) for density is $U_{rc}(\rho_i) = 0.002$.

### 2.3. *Excess molar enthalpy measurements*

The measurements were carried out at 298.15 K using a standard Tian-Calvet micro-calorimeter, self-manufactured, with a temperature stability of 0.01 K. The mixing cell, also designed by us, is made of aluminium and has a small gas phase (< 2 %). Further details on the mixing process and on the calibration of the apparatus have been given previously [17,18]. We remark that the micro-calorimeter was calibrated electrically at 298.15 K and tested performing measurements for the (benzene + cyclohexane) mixture at the same temperature [19,20] (see Table S2 and Figure S2). The observed deviations are ≤ 1.2 % and the estimated maximum relative combined expanded uncertainty (0.95 level of confidence) for $H_m^E$ is 0.03.

## 3. Experimental results

Our results on $V_m^E$ and $H_m^E$ are listed in Table 3 (Figures 1 and 2). Density values are collected in Table S3. The data of the excess molar properties were fitted to the equation:

$$F_{\mathrm{m}}^{\mathrm{E}} = x_1(1-x_1)\sum_{\mathrm{i}=0}^{k-1} A_{\mathrm{i}}(2x_1-1)^{\mathrm{i}} \tag{1}$$

where $F_{\mathrm{m}}^{\mathrm{E}} = V_{\mathrm{m}}^{\mathrm{E}}; H_{\mathrm{m}}^{\mathrm{E}}$ or $U_{V\mathrm{m}}^{\mathrm{E}}$ (see below). The number of coefficients $k$ used in equation (1) for each mixture was determined by applying an F-test [21] at the 99.5 % confidence level. The coefficients are listed in Table 4, together with the corresponding standard deviations for $F_{\mathrm{m}}^{\mathrm{E}}$, $\sigma\left(F_{\mathrm{m}}^{\mathrm{E}}\right)$, calculated using the equation:

$$\sigma\left(F_{\mathrm{m}}^{\mathrm{E}}\right) = \left[\frac{1}{N-k}\sum\left(F_{\mathrm{m,exp}}^{\mathrm{E}} - F_{\mathrm{m,calc}}^{\mathrm{E}}\right)^2\right]^{1/2} \tag{2}$$

where $N$ is the number of data points. From our $H_{\mathrm{m}}^{\mathrm{E}}$ measurements over the whole concentration range, it is possible to determine the excess partial molar enthalpy at infinite dilution of 2-undecane in the heptane mixture. The value is 4.1 kJ mol$^{-1}$, in good agreement with the experimental result (4.4 kJ mol$^{-1}$) [22].

We have also determined values of $U_{V\mathrm{m}}^{\mathrm{E}}$ from the equation [23]:

$$U_{V\mathrm{m}}^{\mathrm{E}} = H_{\mathrm{m}}^{\mathrm{E}} - T\frac{\alpha_p}{\kappa_T}V_{\mathrm{m}}^{\mathrm{E}} \tag{3}$$

In this equation, $\alpha_p$ and $\kappa_T$ stand for, respectively, the isobaric thermal expansion coefficient and the coefficient of isothermal compressibility of the solution. The $T\frac{\alpha_p}{\kappa_T}V_{\mathrm{m}}^{\mathrm{E}}$ term is the so-called contribution from the equation of state (eos) term to $H_{\mathrm{m}}^{\mathrm{E}}$. If the needed data are not available, the $\alpha_p$ and $\kappa_T$ values can be calculated assuming that the mixtures behave ideally with regards to these properties. This procedure is commonly acceptable and has been applied in the present research. Thus, the $\alpha_p$ and $\kappa_T$ values have been determined from the expression:

$$M^{\mathrm{id}} = \phi_1 M_1 + \phi_2 M_2 \tag{4}$$

where $M_{\mathrm{i}} = \alpha_{p\mathrm{i}}$, or $\kappa_{T\mathrm{i}}$ and, $\phi_i = x_i V_{\mathrm{m},i} / (x_1 V_{\mathrm{m},1} + x_2 V_{\mathrm{m},2})$. For 2-undecanone, the values of $V_{\mathrm{mi}}$, $\alpha_{p\mathrm{i}}$ and $\kappa_{T\mathrm{i}}$ are, respectively, 207.10 cm$^3$ mol$^{-1}$; 0.926×10$^{-3}$ K$^{-1}$ (this work) and 816×10$^{-12}$ TPa$^{-1}$ [24]. For other compounds, the values of these quantities were taken from [2]. The $U_{V\mathrm{m}}^{\mathrm{E}}$ data (Figure S3) were fitted using equation (1). The coefficients together with the standard deviations are also listed in Table 4.

**TABLE 3**

Excess molar volumes, $V_m^E$, and enthalpies, $H_m^E$, at temperature $T$ and 95 kPa *vs.* the mole fraction of the ketone, $x_1$, for 2-undecanone (1) + *n*-alkane (2) mixtures.

| $x_1$ | $V_m^E$ /cm$^3$ mol$^{-1}$ | | | $x_1$ | $H_m^E$ /J mol$^{-1}$ |
|---|---|---|---|---|---|
| | $T$/K = 293.15 | $T$/K = 298.15 | $T$/K = 303.15 | | $T$/K = 298.15 |
| 2-undecanone (1) + heptane (2) | | | | | |
| 0.0119 | 0.0027 | 0.0015 | 0.0002 | 0.0655 | 248 |
| 0.0301 | 0.0057 | 0.0032 | 0.0003 | 0.1020 | 347 |
| 0.0453 | 0.0023 | –0.0017 | –0.0038 | 0.1587 | 458 |
| 0.0713 | –0.0031 | –0.0093 | –0.0120 | 0.1987 | 523 |
| 0.0967 | –0.0198 | –0.0243 | –0.0317 | 0.3020 | 604 |
| 0.1445 | –0.0481 | –0.0587 | –0.0689 | 0.3488 | 616 |
| 0.1953 | –0.0653 | –0.0776 | –0.0917 | 0.4254 | 625 |
| 0.2537 | –0.0934 | –0.1074 | –0.1238 | 0.4954 | 604 |
| 0.2994 | –0.1199 | –0.1336 | –0.1501 | 0.6078 | 494 |
| 0.3500 | –0.1327 | –0.1498 | –0.1684 | 0.6996 | 425 |
| 0.4006 | –0.1531 | –0.1714 | –0.1911 | 0.7993 | 304 |
| 0.4555 | –0.1725 | –0.1907 | –0.2108 | 0.8986 | 139 |
| 0.4993 | –0.1778 | –0.1968 | –0.2126 | | |
| 0.5533 | –0.1796 | –0.1968 | –0.2152 | | |
| 0.6068 | –0.1773 | –0.1942 | –0.2105 | | |
| 0.6590 | –0.1724 | –0.1878 | –0.2030 | | |
| 0.7386 | –0.1596 | –0.1672 | –0.1807 | | |
| 0.8114 | –0.1374 | –0.1490 | –0.1593 | | |
| 0.8667 | –0.1098 | –0.1181 | –0.1276 | | |
| 0.9120 | –0.0895 | –0.0954 | –0.1016 | | |
| 2-undecanone (1) + octane (2) | | | | | |
| 0.0452 | 0.0317 | 0.0324 | 0.0326 | 0.0618 | 240 |
| 0.0929 | 0.0522 | 0.0527 | 0.0526 | 0.1119 | 386 |
| 0.1347 | 0.0595 | 0.0584 | 0.0577 | 0.2066 | 545 |
| 0.1971 | 0.0607 | 0.0566 | 0.0533 | 0.3137 | 637 |
| 0.2448 | 0.0511 | 0.0465 | 0.0415 | 0.3867 | 678 |
| 0.2953 | 0.0384 | 0.0341 | 0.0300 | 0.5173 | 637 |
| 0.3464 | 0.0223 | 0.0167 | 0.0122 | 0.6178 | 547 |
| 0.4014 | 0.0119 | 0.0050 | 0.0001 | 0.7036 | 448 |
| 0.4521 | 0.0045 | –0.0039 | –0.0108 | 0.7990 | 311 |
| 0.4993 | –0.0058 | –0.0145 | –0.0209 | 0.8991 | 165 |
| 0.5542 | –0.0134 | –0.0227 | –0.0288 | | |
| 0.6021 | –0.0276 | –0.0340 | –0.0394 | | |
| 0.6534 | –0.0375 | –0.0444 | –0.0497 | | |
| 0.7133 | –0.0446 | –0.0520 | –0.0560 | | |
| 0.7550 | –0.0492 | –0.0550 | –0.0598 | | |
| 0.8077 | –0.0521 | –0.0550 | –0.0581 | | |
| 0.8558 | –0.0527 | –0.0557 | –0.0564 | | |
| 0.9052 | –0.0496 | –0.0524 | –0.0522 | | |
| 2-undecanone (1) + decane (2) | | | | | |
| 0.0359 | 0.0557 | 0.0567 | 0.0590 | 0.0822 | 318 |
| 0.0941 | 0.1249 | 0.1260 | 0.1297 | 0.1129 | 394 |
| 0.2064 | 0.1855 | 0.1902 | 0.1960 | 0.1676 | 522 |
| 0.2482 | 0.2025 | 0.2061 | 0.2118 | 0.2222 | 600 |
| 0.3308 | 0.2141 | 0.2178 | 0.2241 | 0.3012 | 691 |
| 0.4053 | 0.2080 | 0.2131 | 0.2159 | 0.3955 | 732 |
| 0.4301 | 0.2046 | 0.2101 | 0.2122 | 0.5068 | 701 |
| 0.4542 | 0.1998 | 0.2072 | 0.2074 | 0.6106 | 633 |
| 0.4913 | 0.1942 | 0.1986 | 0.2032 | 0.7001 | 529 |
| 0.5383 | 0.1840 | 0.1869 | 0.1916 | 0.8000 | 382 |
| 0.5867 | 0.1733 | 0.1748 | 0.1797 | 0.8984 | 201 |
| 0.6934 | 0.1319 | 0.1340 | 0.1386 | | |
| 0.8155 | 0.0779 | 0.0788 | 0.0851 | | |
| 0.8664 | 0.0572 | 0.0560 | 0.0610 | | |
| 0.8972 | 0.0420 | 0.0432 | 0.0465 | | |
| 0.9703 | 0.0117 | 0.0113 | 0.0145 | | |
| 2-undecanone (1) + dodecane (2) | | | | | |
| 0.0558 | 0.0946 | 0.0968 | 0.0996 | 0.0820 | 317 |
| 0.1032 | 0.1571 | 0.1611 | 0.1674 | 0.1313 | 460 |
| 0.1540 | 0.2116 | 0.2153 | 0.2204 | 0.2036 | 605 |
| 0.2022 | 0.2479 | 0.2532 | 0.2591 | 0.2678 | 695 |
| 0.2511 | 0.2743 | 0.2792 | 0.2845 | 0.3024 | 733 |
| 0.3050 | 0.2975 | 0.3032 | 0.3107 | 0.4136 | 782 |
| 0.3524 | 0.3120 | 0.3186 | 0.3260 | 0.4916 | 777 |
| 0.4175 | 0.3152 | 0.3220 | 0.3302 | 0.6108 | 695 |
| 0.4524 | 0.3102 | 0.3173 | 0.3250 | 0.7085 | 579 |
| 0.5004 | 0.3025 | 0.3106 | 0.3191 | 0.8066 | 409 |
| 0.5682 | 0.2798 | 0.2866 | 0.2953 | 0.8526 | 328 |
| 0.6464 | 0.2481 | 0.2552 | 0.2605 | 0.8973 | 230 |
| 0.7111 | 0.2180 | 0.2241 | 0.2304 | | |
| 0.7845 | 0.1742 | 0.1772 | 0.1817 | | |
| 0.8498 | 0.1211 | 0.1248 | 0.1281 | | |
| 0.8990 | 0.0821 | 0.0840 | 0.0862 | | |
| 0.9498 | 0.0331 | 0.0358 | 0.0374 | | |
| 2-undecanone (1) + tetradecane (2) | | | | | |
| 0.0611 | 0.1287 | 0.1343 | 0.1321 | 0.0611 | 253 |
| 0.0991 | 0.1887 | 0.1906 | 0.1883 | 0.1302 | 475 |
| 0.1503 | 0.2448 | 0.2505 | 0.2521 | 0.1423 | 504 |
| 0.2028 | 0.2994 | 0.3053 | 0.3045 | 0.1907 | 619 |
| 0.2511 | 0.3366 | 0.3436 | 0.3454 | 0.2224 | 672 |
| 0.2982 | 0.3575 | 0.3715 | 0.3793 | 0.3170 | 803 |
| 0.3505 | 0.3805 | 0.3884 | 0.3905 | 0.4162 | 850 |
| 0.3996 | 0.3861 | 0.3952 | 0.4016 | 0.4981 | 849 |
| 0.4508 | 0.3882 | 0.3970 | 0.4034 | 0.6034 | 789 |
| 0.5140 | 0.3807 | 0.3896 | 0.3972 | 0.7064 | 658 |
| 0.5476 | 0.3739 | 0.3831 | 0.3903 | 0.8010 | 480 |
| 0.5996 | 0.3519 | 0.3597 | 0.3655 | 0.8547 | 372 |
| 0.6492 | 0.3221 | 0.3305 | 0.3355 | 0.8994 | 270 |
| 0.7402 | 0.2661 | 0.2722 | 0.2781 | | |
| 0.8465 | 0.1683 | 0.1743 | 0.1807 | | |
| 0.8992 | 0.1137 | 0.1168 | 0.1176 | | |
| 0.9507 | 0.0513 | 0.0542 | 0.0542 | | |

[a]The standard uncertainties, $u$, are: $u(T)$ = 0.01 K; $u(p)$ = 10 kPa; $u(x_1)$ = 0.0005; $u(V_\mathrm{m}^\mathrm{E})$ = 0.010$\left|V_\mathrm{m}^\mathrm{E}\right|_\mathrm{max}$ + 0.005 cm$^3$ mol$^{-1}$. For $H_\mathrm{m}^\mathrm{E}$, the relative combined expanded uncertainty (0.95 level of confidence) is $U_\mathrm{rc}\left(H_\mathrm{m}^\mathrm{E}\right)$ = 0.030.

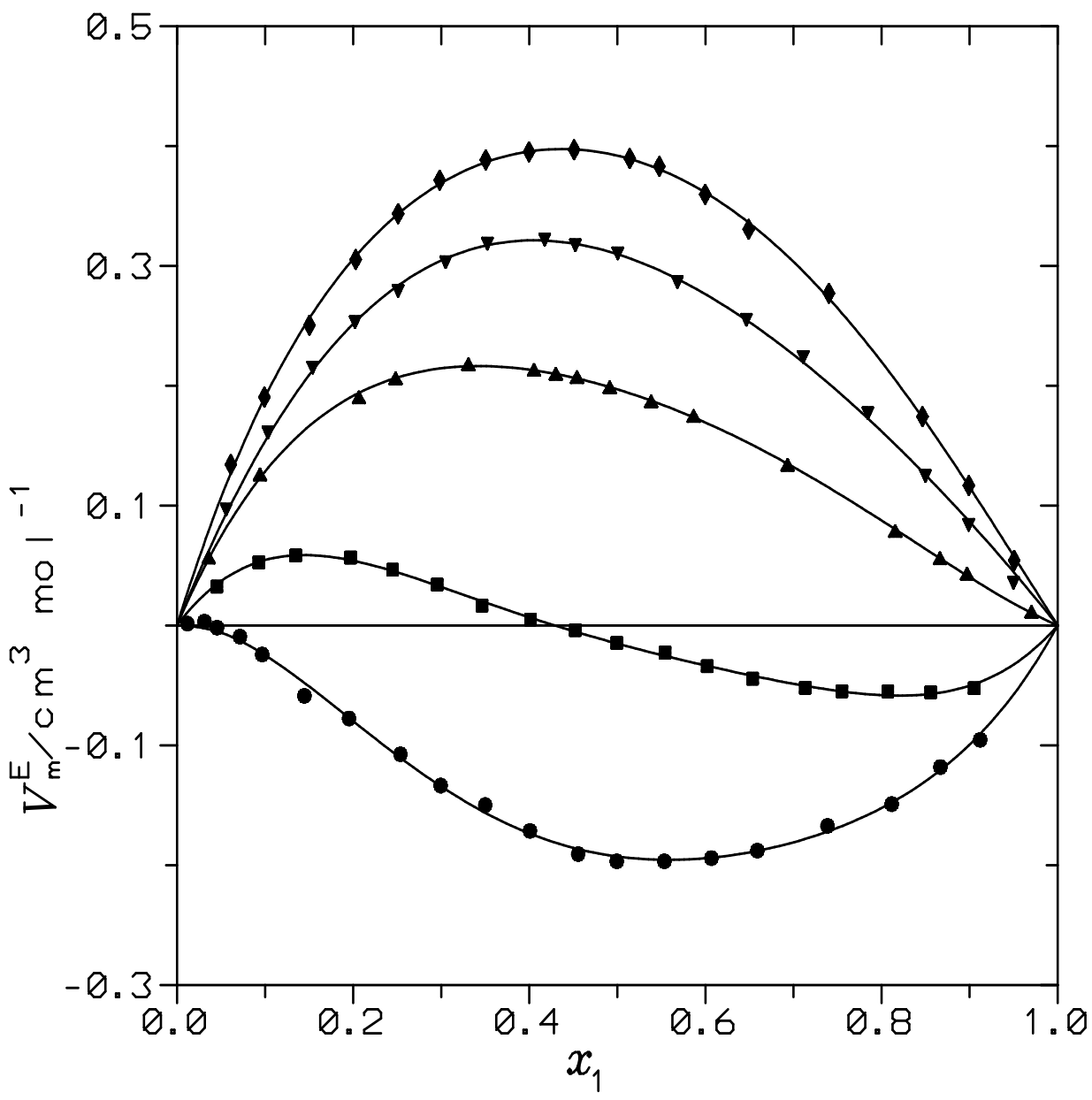

**Figure 1** $V_{\mathrm{m}}^{\mathrm{E}}$ of 2-undecanone (1) + *n*-alkane (2) mixtures at 298.15 K and 95 kPa. Points, experimental results (this work): (●), heptane; (■), octane; (▲), decane; (▼), dodecane. (♦); tetradecane. Solid lines, calculations with equation (1) using coefficients listed in Table 4.

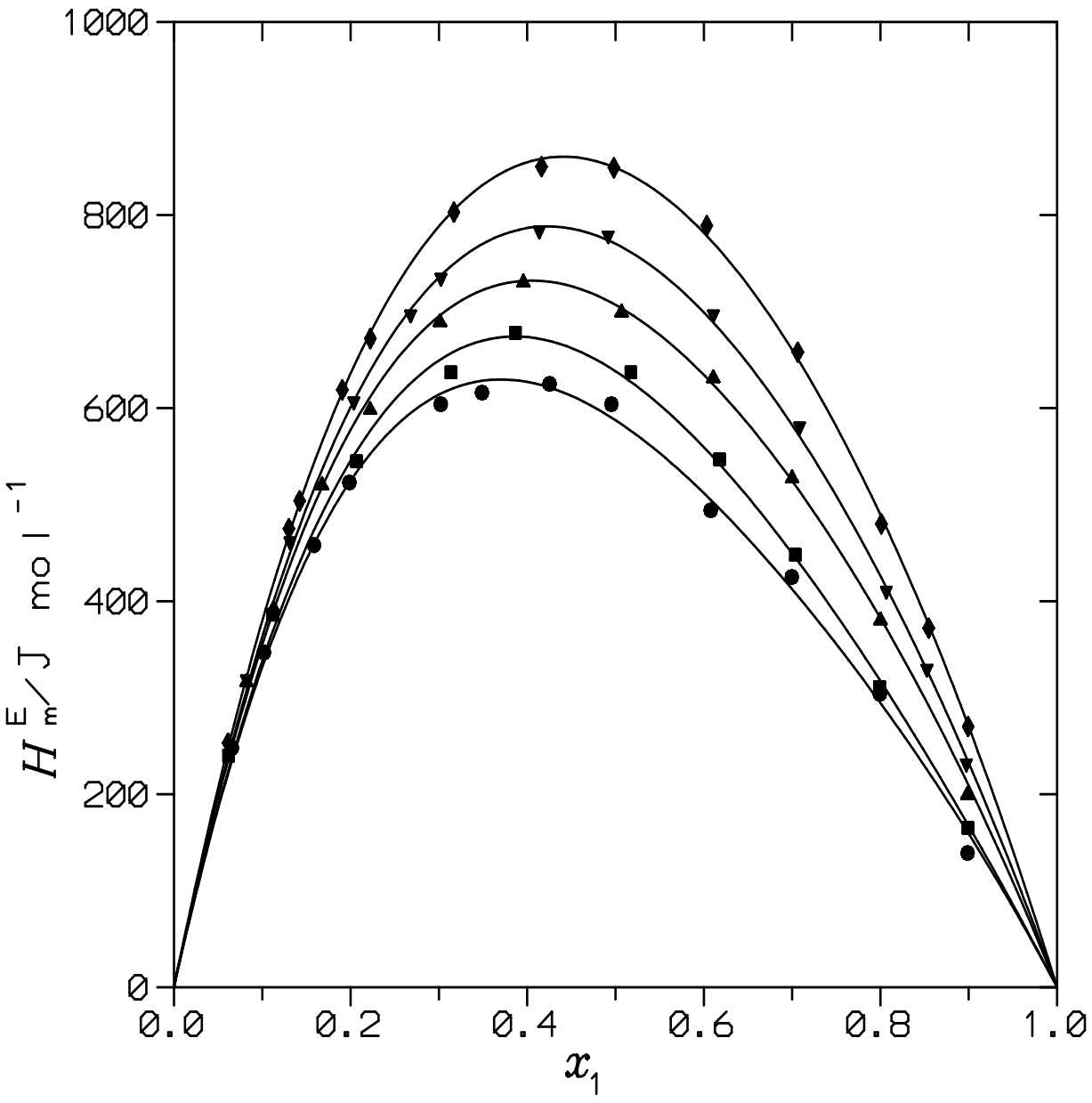

**Figure 2** $H_{\mathrm{m}}^{\mathrm{E}}$ of 2-undecanone (1) + *n*-alkane (2) mixtures at 298.15 K and 95 kPa. Points, experimental results (this work): (●), heptane; (■), octane; (▲), decane; (▼), dodecane. (♦); tetradecane. Solid lines, calculations with equation (1) using coefficients listed in Table 4.

**TABLE 4**

Coefficients $A_i$ and standard deviations, $\sigma(F_m^E)$ (equation (2)) for the representation of the excess molar functions $F_m^E$ (= $V_m^E$ ; $H_m^E$ ; $U_{Vm}^E$ ) at temperature $T$ and 95 kPa for 2-undecanone (1) + $n$-alkane (2) systems by equation (1).

| $n$-alkane | $F_m^E$ | $T$/K | $A_0$ | $A_1$ | $A_2$ | $A_3$ | $\sigma(F_m^E)$ [a] |
|---|---|---|---|---|---|---|---|
| $n$-$C_7$ | $V_m^E$ | 293.15 | –0.699 | –0.218 | 0.120 | –0.406 | 0.004 |
| | | 298.15 | –0.7711 | –0.197 | 0.123 | –0.502 | 0.004 |
| | | 303.15 | –0.845 | –0.166 | 0.118 | –0.556 | 0.004 |
| | $H_m^E$ | 298.15 | 2350 | –1200 | 586 | | 13 |
| | $U_{Vm}^E$ | 298.15 | 2583 | –1067 | 563 | | 2 |
| $n$-$C_8$ | $V_m^E$ | 293.15 | –0.028 | –0.393 | 0.103 | –0.505 | 0.002 |
| | | 298.15 | –0.061 | –0.398 | 0.133 | –0.517 | 0.002 |
| | | 303.15 | –0.086 | –0.399 | 0.159 | –0.513 | 0.002 |
| | $H_m^E$ | 298.15 | 2553 | –1189 | 398 | | 13 |
| | $U_{Vm}^E$ | 298.15 | 2571 | –1027 | 360 | | 3 |
| $n$-$C_{10}$ | $V_m^E$ | 293.15 | 0.771 | –0.424 | 0.247 | –0.280 | 0.002 |
| | | 298.15 | 0.788 | –0.444 | 0.294 | –0.255 | 0.002 |
| | | 303.15 | 0.802 | 0.444 | 0.294 | –0.268 | 0.002 |
| | $H_m^E$ | 298.15 | 2828 | –1017 | 477 | | 9 |
| | $U_{Vm}^E$ | 298.15 | 2587 | –874 | 416 | | 1 |
| $n$-$C_{12}$ | $V_m^E$ | 293.15 | 1.212 | –0.414 | 0.145 | –0.143 | 0.003 |
| | | 298.15 | 1.239 | –0.467 | 0.149 | | 0.004 |
| | | 303.15 | 1.272 | –0.413 | 0.153 | –0.181 | 0.003 |
| | $H_m^E$ | 298.15 | 3081 | –911 | 357 | | 7 |
| | $U_{Vm}^E$ | 298.15 | 2696 | –794 | 317 | | 1 |
| $n$-$C_{14}$ | $V_m^E$ | 293.15 | 1.531 | –0.344 | 0.188 | –0.287 | 0.003 |
| | | 298.15 | 1.568 | –0.346 | 0.208 | –0.279 | 0.004 |
| | | 303.15 | 1.560 | –0.327 | 0.167 | –0.276 | 0.004 |
| | $H_m^E$ | 298.15 | 3396 | –762 | 336 | | 6 |
| | $U_{Vm}^E$ | 298.15 | 2899 | –651 | 273 | | 1 |

[a] $\sigma(F_m^E) = \left[ \frac{1}{N-k} \sum \left( F_{m,exp}^E - F_{m,calc}^E \right)^2 \right]^{1/2}$ ; units are: J mol$^{-1}$ ( $F_m^E = H_m^E ; U_{Vm}^E$ ), or cm$^3$ mol$^{-1}$ ( $F_m^E = V_m^E$ ).

## 4. Discussion

Below, values of the excess molar properties are provided at equimolar composition and 298.15 K.

### *4.1. Excess molar enthalpy results*

(2-Undecanone + $n$-alkane) systems are characterized by positive values of $H_m^E$ (Table 3, Figure 2), and this means that interactions between like molecules are prevalent. On the other hand, $H_m^E$ values and $n$ (the number of C atoms in the $n$-alkane) increase in line (Figure 3), which is the normal behaviour. In fact, the same trend is observed for other (2-alkanone + $n$-alkane) systems (Figure 3). For example, $H_m^E$ (2-butanone)/J mol$^{-1}$ = 1338 ($n$ = 7) [25]; 1408 ($n$ = 8) [25];

1545 ($n$ = 10) [25]; 1658 ($n$ = 12) [26]. On the other hand, for systems with a given alkane, $H_{\mathrm{m}}^{\mathrm{E}}$ decreases when the 2-alkanone size increases: $H_{\mathrm{m}}^{\mathrm{E}}$ (heptane)/J mol$^{-1}$ = 1338 (2-butanone) [25]; 1055 (2-hexanone) [27]; 587 (2-undecanone) (see Figure 3). This can be due to the corresponding weakening of the interactions between ketone molecules as it is supported by the relative variation of the partial excess molar enthalpy at infinite dilution of the 2-alkanone ($H_{\mathrm{m1}}^{\mathrm{E},\infty}$) in mixtures with heptane since $H_{\mathrm{m1}}^{\mathrm{E},\infty}$/kJ mol$^{-1}$ = 7.5 (2-butanone) [25]; 6.6 (2-hexanone) [27]; 4.1 (2-undecanone). The weakening of ketone-ketone interactions may be ascribed to the corresponding weakening of dipolar interactions, which can be investigated by means of effective dipole moment, $\bar{\mu}_{\mathrm{i}}$, defined by [23,28]:

$$\bar{\mu}_{\mathrm{i}} = \left[\frac{\mu_{\mathrm{i}}^2 N_{\mathrm{A}}}{4\pi\varepsilon_{\mathrm{o}} V_{\mathrm{mi}} k_{\mathrm{B}} T}\right]^{1/2} \qquad (5)$$

where all the symbols have the usual meaning. For 2-alkanones, we have: $\bar{\mu}_{\mathrm{i}}$ = 1.28 (propanone); 1.12 (2-butanone); 0.91 (2-hexanone); 0.82 (2-octanone) [2]; 0.72 (2-undecanone, value obtained using $\mu_{\mathrm{i}}$ = 2.71 D [29]). This set of results reveals that dipolar interactions between ketone molecules become weaker when the alkanone size increases.

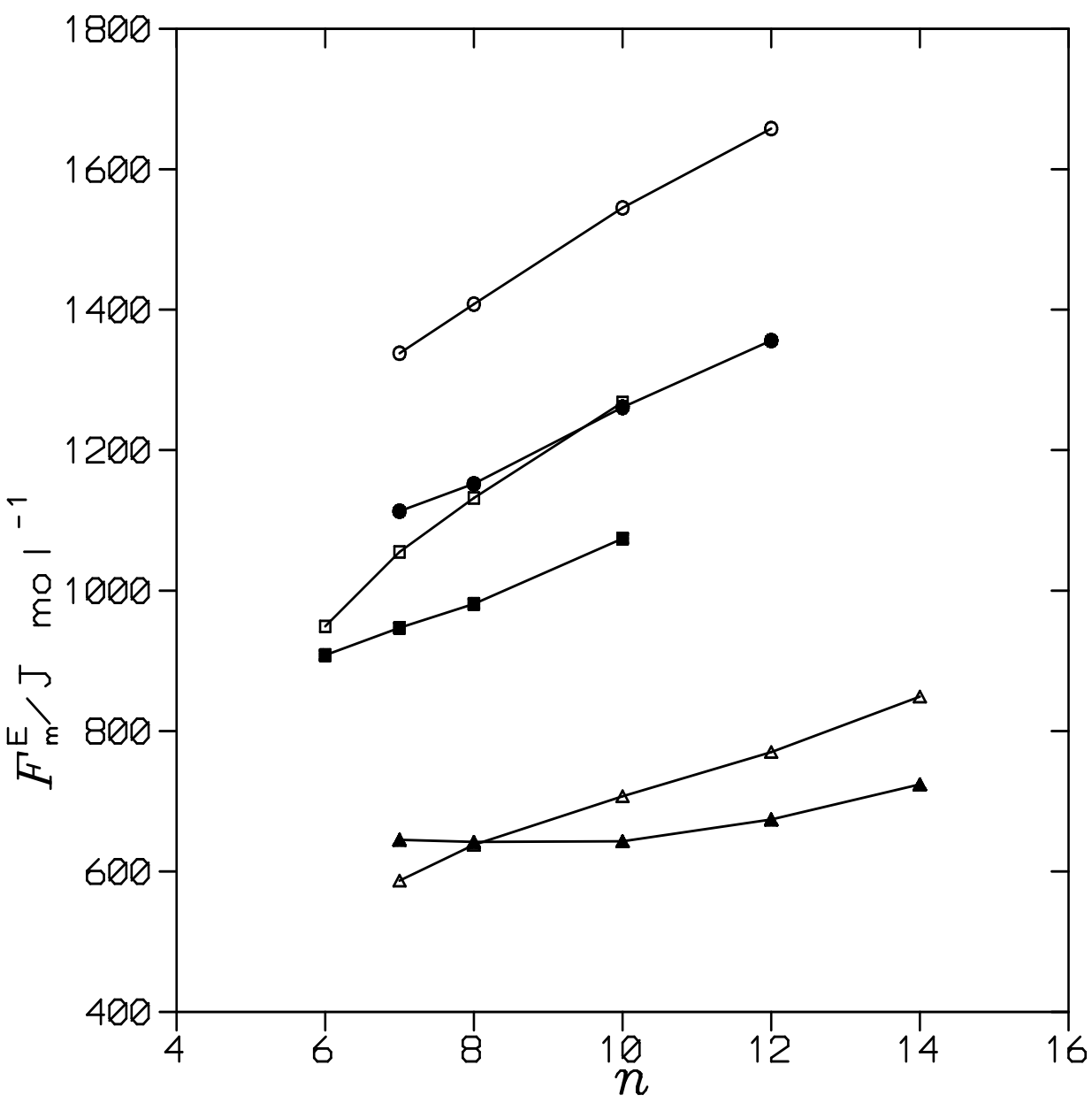

**Figure 3** Excess molar functions $F_{\mathrm{m}}^{\mathrm{E}}$ for 2-alkanone (1) + $n$-alkane (2) mixtures at equimolar compositions, 298.15 K and atmospheric pressure *vs.* $n$, the number of C atoms in the alkane. Open symbols, $F_{\mathrm{m}}^{\mathrm{E}} = H_{\mathrm{m}}^{\mathrm{E}}$ : (O), 2-butanone [25,26]; (□), 2-hexanone [27]; (Δ), 2-undecanone (this work). Full symbols, $F_{\mathrm{m}}^{\mathrm{E}} = U_{V\mathrm{m}}^{\mathrm{E}}$ : (●), 2-butanone; (■), 2-hexanone; (▲), 2-undecanone. Lines are for the aid of the eye.

### 4.2. *Volumetric data*

Firstly, we remark that the $V_{\mathrm{m}}^{\mathrm{E}}$ result for (2-undecanone + heptane) system is negative ( $-0.1922$ cm$^3$ mol$^{-1}$) and this clearly indicates that $V_{\mathrm{m}}^{\mathrm{E}}$ is mainly determined by structural effects since the corresponding $H_{\mathrm{m}}^{\mathrm{E}}$ value is positive (587 J mol$^{-1}$). Such effects are also present, although to a lesser extent, in the solution with $n$ = 8 (Figure 1). The mentioned effects can be due to the different size of the mixture compounds and to free volume effects. Thus, $V_{\mathrm{mi}}$ /cm$^3$ mol$^{-1}$ = 207.01 (2-undecanone); 147.45 (heptane) and $\alpha_{p\mathrm{i}}$ /10$^{-3}$ K$^{-1}$ = 0.926 (2-undecanone) and 1.256 (heptane) [2]. We note that $V_{\mathrm{m}}^{\mathrm{E}}$ increases with $n$ (Figure 1) and this can be ascribed, in part, to decreasing free volume effects but also to an increased interactional contribution to $V_{\mathrm{m}}^{\mathrm{E}}$, as it is supported by the present $H_{\mathrm{m}}^{\mathrm{E}}$ data. Accordingly with this statement, the $A_p = \left(\frac{\Delta V_{\mathrm{m}}^{\mathrm{E}}}{\Delta T}\right)_p$ values change from a negative value for the system with $n$ = 7 ($-3.6\times10^{-3}$ cm$^3$ mol$^{-1}$ K$^{-1}$) to positive results for the solutions with the longer $n$-alkanes ($1.5\times10^{-3}$ cm$^3$ mol$^{-1}$ K$^{-1}$ for $n$ = 12).

For mixtures with other 2-alkanones, the values of $V_{\mathrm{m}}^{\mathrm{E}}$ also increase with $n$ (Table S4). For 2-butanone systems, $V_{\mathrm{m}}^{\mathrm{E}}$ /cm$^3$ mol$^{-1}$ = 0.803 ($n$ = 7) [30]; 0.887 ($n$ = 8) [31]; 0.952 ($n$ = 10) [30]; 0.996 ($n$ = 12) [30]. These values are much higher than those of the 2-undecanone mixtures, and reveal that they are largely determined by much stronger interactional effects, since such effects become weaker when the 2-alkanone increases. The values of $A_p$ are consistent with this analysis. Thus, $A_p$ (heptane)/cm$^3$ mol$^{-1}$ K$^{-1}$ = $8.5\times10^{-3}$ (2-butanone); $4.9\times10^{-3}$ (2-hexanone) [30,32-34].

### 4.3. *Isochoric excess molar internal energies*

Results on $U_{V\mathrm{m}}^{\mathrm{E}}$ for mixtures with 2-alkanones are listed in Table S4 (Figure 3 and S3). For heptane systems, the isochoric partial excess molar internal energies at infinite dilution of the alkanone, ($U_{V\mathrm{m1}}^{\mathrm{E},\infty}$/kJ mol$^{-1}$), decrease when the ketone size increases: 6.1 (2-butanone); 5.7 (2-hexanone); 4.2 (2-undecanone), which reinforces that interactions between alkanone molecules become weaker at the mentioned condition. On the other hand, we note that for systems with 2-butanone or 2-hexanone, the $U_{V\mathrm{m}}^{\mathrm{E}}$ results are lower than the corresponding $H_{\mathrm{m}}^{\mathrm{E}}$ results and that they increase in line with $n$ (Figure 3). A different trend is observed for the $U_{V\mathrm{m}}^{\mathrm{E}}$ values of solutions involving 2-undecanone: (i) for the mixture with $n$ = 7, $H_{\mathrm{m}}^{\mathrm{E}} < U_{V\mathrm{m}}^{\mathrm{E}}$, which remarks the relevance of the eos contribution to $H_{\mathrm{m}}^{\mathrm{E}}$; (ii) $U_{V\mathrm{m}}^{\mathrm{E}}$ values remain practically constant in the range $n$ = 7-10, and then slightly increase. This can be explained in terms of a possible folding of 2-undecanone in the investigated systems. We have demonstrated previously that in mixtures formed by a cyclic molecule (benzene, cyclohexane, $C_6H_5X$ (X = Cl,Br,I)) [35-37], $U_{V\mathrm{m}}^{\mathrm{E}}$ is the result of two contributions. The first one arises from the poorer ability of longer $n$-alkanes to disrupt interactions between the mentioned cyclic molecules and leads to decreased $U_{V\mathrm{m}}^{\mathrm{E}}$ values. The second contribution is consequence of the breaking of the correlations of molecular orientations (CMO), a local order characteristic of longer $n$-alkanes and leads to increased values of $U_{V\mathrm{m}}^{\mathrm{E}}$. Thus, if one assumes that 2-undecanone adopts quasi-cyclic structures in the present systems, both contributions are counterbalanced for $n$ = 7-10, while the second one slightly dominates for $n$ =12,14.

As in a previous work [2], we investigate now orientational effects in 2-undecanone systems by means of the Flory model. The main hypotheses and equations of the theory can be found elsewhere [5,38]. Now, we merely remark that random mixing is a basic assumption of the model. In order to conduct a more detailed analysis, we have applied shortly the model using both $H_{\mathrm{m}}^{\mathrm{E}}$ and $U_{V\mathrm{m}}^{\mathrm{E}}$ data. For $H_{\mathrm{m}}^{\mathrm{E}}$, the Flory interaction parameters ($X_{12}$ /J cm$^{-3}$) are: 16.14 ($n$ = 7); 16.47 ($n$ = 8); 16.74 ($n$ = 10); 17.15 ($n$ = 12); 18.11 ($n$ = 14). The corresponding standard relative deviations for $H_{\mathrm{m}}^{\mathrm{E}}$ defined by:

$$\sigma_{\mathrm{r}}\left(H_{\mathrm{m}}^{\mathrm{E}}\right)=\left[\frac{1}{N}\sum\left(\frac{H_{\mathrm{m,exp}}^{\mathrm{E}}-H_{\mathrm{m,calc}}^{\mathrm{E}}}{H_{\mathrm{m,exp}}^{\mathrm{E}}}\right)^{2}\right]^{1/2} \tag{6}$$

are: 0.215 ($n$ = 7); 0.206 ($n$ = 8); 0.186 ($n$ = 10); 0.179 ($n$ = 12); 0.175 ($n$ = 14). For $U_{V\mathrm{m}}^{\mathrm{E}}$ values, ($X_{12}$ /J cm$^{-3}$) = 17.66 ($n$ = 7); 16.57 ($n$ = 8); 15.22 ($n$ = 10); 15.02 ($n$ = 12); 15.45 ($n$ = 14); and, in the same order, $\sigma_{\mathrm{r}}\left(U_{V\mathrm{m}}^{\mathrm{E}}\right)$ = 0.130; 0.146; 0.150; 0.162; 0.155. These results indicate that orientational effects are largely present in the solutions. It is to be noted that $\sigma_{\mathrm{r}}\left(U_{V\mathrm{m}}^{\mathrm{E}}\right)$ < $\sigma_{\mathrm{r}}\left(H_{\mathrm{m}}^{\mathrm{E}}\right)$, which newly underlines the importance of structural effects on $H_{\mathrm{m}}^{\mathrm{E}}$.

*4.4. Comparison with mixtures containing n-alkanoates*

For the sake of comparison, we have selected mixtures with methyl $n$-decanoate or $n$-dodecanoate, which are esters of similar size to 2-undecanone. Firstly, we note that for solutions with a given alkane, $H_{\mathrm{m}}^{\mathrm{E}}$ (ester) < $H_{\mathrm{m}}^{\mathrm{E}}$ (2-undecanone) (Figure 4). For example, $H_{\mathrm{m}}^{\mathrm{E}}$ (heptane)/J mol$^{-1}$ = 550 (methyl $n$-decanoate) [39]; 442 (methyl $n$-dodecanoate) [40]; 587 (2-undecanone), or $H_{\mathrm{m}}^{\mathrm{E}}$ (tetradecane)/J mol$^{-1}$ = 789 (methyl $n$-decanoate) [41]; 849 (2-undecanone). The same trend is observed for the $H_{\text{m1}}^{\text{E},\infty}$ /kJ mol$^{-1}$ values for heptane mixtures: 3.3 (methyl *n*-decanoate [39]); 3.7 (methyl *n*-dodecanoate [40]); 4.1 (2-undecanone). One can conclude that interactions between ester molecules are weaker than between 2-undecanone molecules. This is confirmed by the corresponding values of $U_{V\text{m1}}^{\text{E},\infty}$(heptane)/kJ mol$^{-1}$: 3.1 (methyl *n*-decanoate); 3.0 (methyl *n*-dodecanoate) [11]; 4.2 (2-undecanone). The same trend is observed when solutions involving smaller molecules are considered, at it is clearly shown by the upper critical solution temperatures of the mixtures (2-propanone + octane) (253.0 K [42]), or (methyl ethanoate + octane) (241.7 K [43]). The most important result is the different variation of $U_{V\text{m}}^{\text{E}}(n)$ for the systems under consideration (Figure 4). For solutions with the alkanoates $CH_3(CH_2)_{u-1}COO(CH_2)_{v-1}CH_3$ ($u$ = 8,10; $v$ = 1), $U_{V\text{m}}^{\text{E}}(n)$ shows a minimum at $n$ = 9, which has been explained by the folding of these alkanoates [11]. Thus, for $n$ =5-9, the contribution to $U_{V\text{m}}^{\text{E}}$ from the poorer ability of longer alkanes to break ester-ester interactions dominates; while from $n > 9$, the contribution related to the disruption of the CMO of longer *n*-alkanes predominates. It is important to note that the folding of *n*-alkanoates is also observed in solutions containing shorter esters, e.g. systems with $u = 4$, $v = 1$. In this case, the minimum of $U_{V\text{m}}^{\text{E}}(n)$ is encountered at $n$ = 6-7 [11]. Structural changes that can support the existence of folding in these types of systems can be examined using $V_{\text{m}}^{\text{E}}$ data. For ($CH_3(CH_2)_{u-1}COOCH_3$ + heptane) systems, the slope of $V_{\text{m}}^{\text{E}}(u)$ changes noticeably at $u$ = 6 [11] (Figure 5). This is in agreement with the fact $U_{V\text{m}}^{\text{E}}(n)$ shows a minimum. For ($CH_3CO(CH_2)_{u-1}CH_3$ + heptane) mixtures, the slope change of $V_{\text{m}}^{\text{E}}(u)$ is smoother, indicating that structural changes linked to folding are weaker (Figure 5).

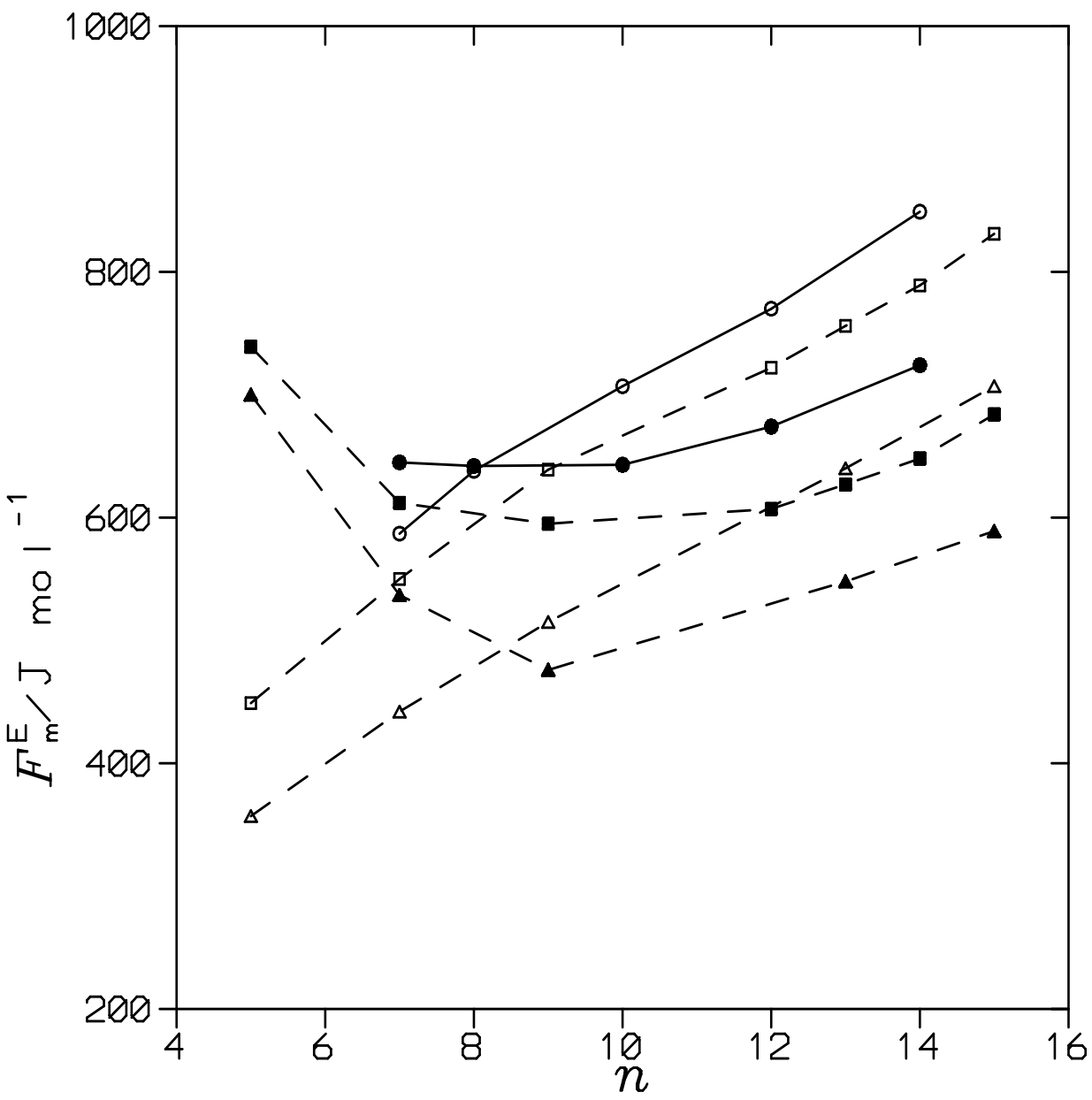

**Figure 4** Excess molar functions $F^{\mathrm{E}}_{\mathrm{m}}$ for polar compound (1) + *n*-alkane (2) mixtures at equimolar compositions, 298.15 K and atmospheric pressure *vs*. *n*, the number of C atoms in the alkane. Open symbols, $F^{\mathrm{E}}_{\mathrm{m}} = H^{\mathrm{E}}_{\mathrm{m}}$ : (O), 2-undecanone; (□), methyl *n*-decanoate; (Δ), methyl *n*-dodecanoate (for source of data see [11] and [63]). Full symbols, $F^{\mathrm{E}}_{\mathrm{m}} = U^{\mathrm{E}}_{V\mathrm{m}}$ : (●), 2-undecanone; (■), methyl *n*-decanoate [11]; (▲), methyl *n*-dodecanoate [11]. Lines are for the aid of the eye.

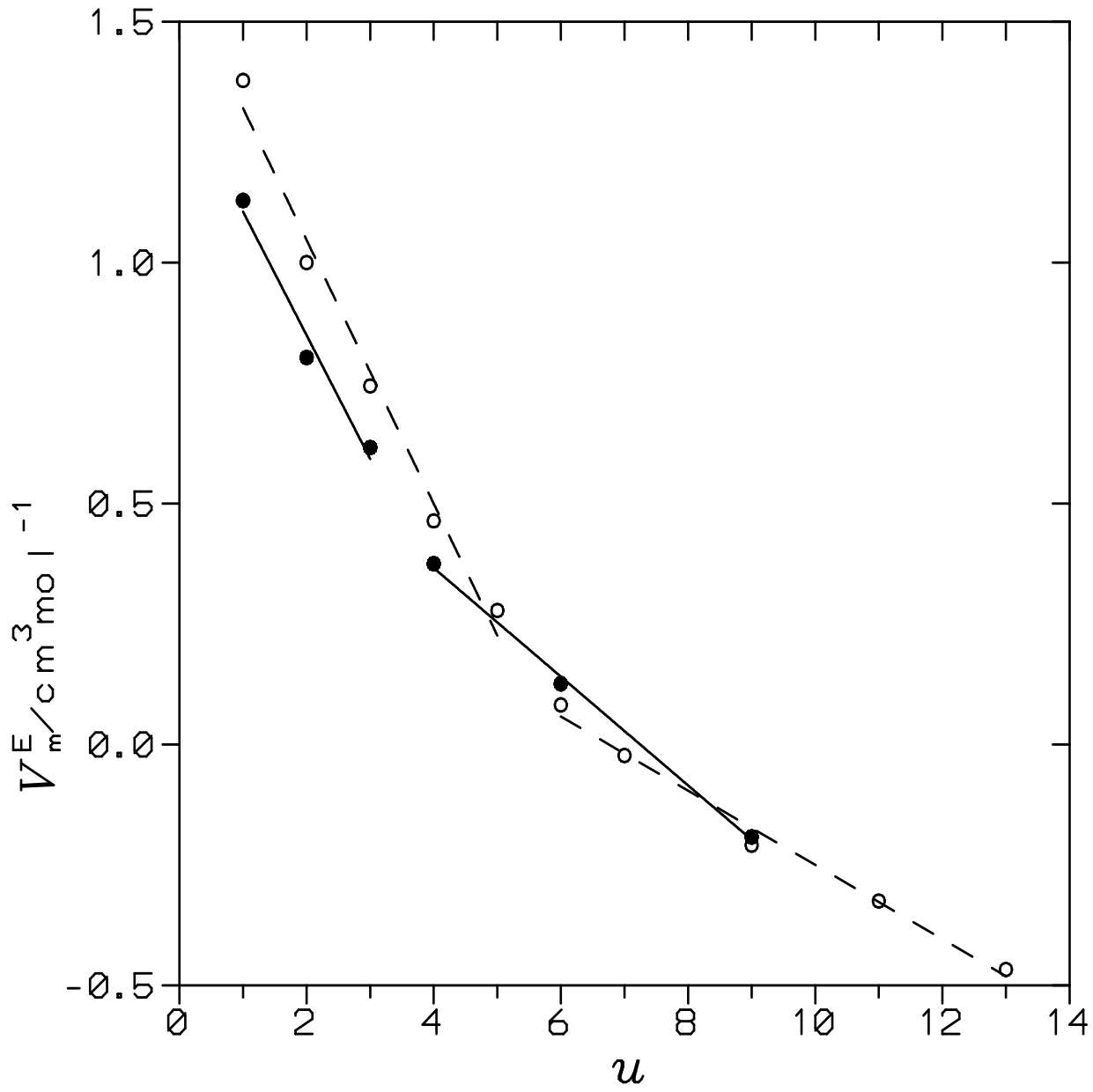

**Figure 5** Excess molar volumes, $V^{\mathrm{E}}_{\mathrm{m}}$ , for $CH_3CO(CH_2)_{u\text{-}1}CH_3$ (1), or + $CH_3(CH_2)_{u\text{-}1}COOCH_3$ (1) + heptane (2) mixtures at equimolar compositions, 298.15 K and atmospheric pressure *vs*. *u*. Full symbols: 2-alkanones [10, 30,31,33]; open symbols, methyl *n*-alkanoates (for source of data, see [11] and [66]). Lines are the results from fittings to straight lines.

## 5. Conclusions

Density and $V_{\rm m}^{\rm E}$ data at (293.15-303.15) K and $H_{\rm m}^{\rm E}$ measurements at 298.15 K have been reported for (2-undecanone + *n*-alkane) mixtures at 95 kPa. Calorimetric data reveal that interactions between like molecules are dominant. The contribution to $V_{\rm m}^{\rm E}$ from structural effects is determinant for systems with *n* =7,8. $H_{\rm m}^{\rm E}$ results increase with *n*, while $U_{V{\rm m}}^{\rm E}$ values are nearly constant for *n* =7-10, and then slightly increases. The $U_{V{\rm m}}^{\rm E}(n)$ variation has been explained in terms of a possible folding of 2-undecanone, which is not present in other 2-alkanones. Folding is more probable in solutions involving *n*-alkanoates.

**Funding**

This work was carried out under Project PID2022-137104NA-I00, funded by MICIN/AEI/10.13039/501100011033/ and by FEDER, UE. F. Pazoki acknowledges the FPI grant PREP2022-000047 from MCIN/AEI/10.13039/501100011033/ and FEDER, UE. The author J. V. Alves-Laurentino would like to acknowledge funding from European Social Fund Plus, Operational Programme of Castilla y León, and from Junta de Castilla y León, through Consejería de Educación.

# THERMODYNAMICS OF 2-UNDECANONE + *N*-ALKANE MIXTURES

## SUPPLEMENTARY MATERIAL

Susana Villa, Fatemeh Pazoki, Juan Antonio González*, Fernando Hevia, Daniel Lozano-Martín, Luis Felipe Sanz, João Victor Alves-Laurentino

G.E.T.E.F., Departamento de Física Aplicada, Facultad de Ciencias, Universidad de Valladolid, Paseo de Belén, 7, 47011 Valladolid, Spain.

*corresponding author, e-mail: jagl@termo.uva.es

**Table S1**

Sample description.

| Chemical name | CAS Number | Source | Initial purity[a] | Purification method |
|---|---|---|---|---|
| cyclohexane[b,c] | 110-82-7 | Fluka | 0.999 | none |
| benzene[b] | 71-43-2 | Sigma -Aldrich | 0.9995 | none |
| heptane[c] | 142-82-5 | Fluka | 0.998 | none |
| isooctane[c] | 540-84-1 | Fluka | 0.999 | none |
| toluene[c] | 108-88-3 | Fluka | > 0.995 | none |
| water HPLC plus[c] | 7732-18-5 | Sigma-Aldrich | | |

[a] Gas chromatography area fraction, certified by the supplier.

[b] Liquids of the reference system used for testing the densitometer Anton Paar DMA 602 and of the Tian-Calvet micro-calorimeter measuring excess molar volumes, and excess molar enthalpies, respectively.

[c] Reference liquids used for the calibration of the densitometer.

**Table S2.** Densities, $\rho$, and excess molar functions, volumes ($V_m^E$) and enthalpies ($H_m^E$), for the cyclohexane (1) + benzene (2) mixture at 298.15 K and 95 kPa, *vs.* $x_1$, the mole fraction of 2-undecanone[a].

| $x_1$ | $\rho$ /g cm$^{-3}$ | ($V_m^E$ /cm$^3$ mol$^{-1}$)[b] | $x_1$ | ($H_m^E$ /J mol$^{-1}$)[c] |
|---|---|---|---|---|
| 0 | 0.87327 | 0 | 0.1028 | 307 |
| 0.0607 | 0.86463 | 0.1465 | 0.2036 | 527 |
| 0.1115 | 0.85772 | 0.2559 | 0.2121 | 548 |
| 0.1644 | 0.85086 | 0.3504 | 0.2138 | 544 |
| 0.2064 | 0.84554 | 0.4233 | 0.3068 | 687 |
| 0.2470 | 0.84062 | 0.4810 | 0.3073 | 683 |
| 0.2899 | 0.83560 | 0.5328 | 0.4115 | 783 |
| 0.3388 | 0.82625 | 0.5823 | 0.5062 | 800 |
| 0.3736 | 0.82028 | 0.6100 | 0.6060 | 756 |
| 0.4304 | 0.81539 | 0.6404 | 0.7128 | 643 |
| 0.4791 | 0.82028 | 0.6510 | 0.8084 | 483 |
| 0.5449 | 0.81539 | 0.6507 | 0.8954 | 291 |
| 0.6039 | 0.80905 | 0.6319 | | |
| 0.6619 | 0.80365 | 0.5942 | | |
| 0.7012 | 0.79859 | 0.5502 | | |
| 0.7632 | 0.79041 | 0.4773 | | |
| 0.7800 | 0.78907 | 0.4598 | | |
| 0.8422 | 0.78443 | 0.3588 | | |
| 0.8864 | 0.78131 | 0.2724 | | |
| 0.9348 | 0.77802 | 0.1644 | | |
| 1 | 0.77382 | 0 | | |

[a] The standard uncertainties, $u$, are: $u(T)$ = 0.01 K; $u(p)$ = 10 kPa; $u(x_1)$ = 0.0005; $u(V_m^E)$= 0.010 + 0.005 cm$^3$ mol$^{-1}$. For density and $H_m^E$, the relative combined expanded uncertainties (0.95 level of confidence) are: $U_{rc}(\rho)$ = 0.002 and $U_{rc}(H_m^E)$ = 0.030, respectively.

[b] Coefficients of Eq. (1): $A_0$ = 2.622; $A_1$ = 0.076; $\sigma(V_m^E)$ [Eq. (2)] = 0.002 cm$^3$ mol$^{-1}$.

[c] Coefficients of Eq. (1): $A_0$ = 3193; $A_1$ = −112; $\sigma(H_m^E)$ [Eq. (2)] = 3.4 J mol$^{-1}$.

**Table S3**

Densities of 2-undecanone (1) + *n*-alkane (2) mixtures at temperature $T$ and 95 kPa[a] *vs.* $x_1$, the mole fraction of 2-undecanone.

| | $\rho$ /g cm$^{-3}$ | | |
|---|---|---|---|
| $x_1$ | $T$/K = 293.15 | $T$/K = 298.15 | $T$/K = 303.15 |
| 2-undecanone (1) + heptane (2) | | | |
| 0.0119 | 0.68619 | 0.68196 | 0.67771 |
| 0.0310 | 0.68994 | 0.68571 | 0.68148 |
| 0.0453 | 0.69273 | 0.68852 | 0.68428 |
| 0.0713 | 0.69773 | 0.69353 | 0.68931 |
| 0.0967 | 0.70255 | 0.69836 | 0.69416 |
| 0.1445 | 0.71137 | 0.70722 | 0.70305 |
| 0.1953 | 0.72035 | 0.71622 | 0.71209 |
| 0.2537 | 0.73029 | 0.72619 | 0.72210 |
| 0.2994 | 0.73779 | 0.73370 | 0.72963 |
| 0.3500 | 0.74575 | 0.74170 | 0.73766 |
| 0.4006 | 0.75346 | 0.74943 | 0.74542 |
| 0.4555 | 0.76151 | 0.75751 | 0.75352 |
| 0.4993 | 0.76767 | 0.76369 | 0.75971 |
| 0.5533 | 0.77501 | 0.77104 | 0.76710 |
| 0.6068 | 0.78200 | 0.77806 | 0.77413 |
| 0.6590 | 0.78858 | 0.78465 | 0.78075 |
| 0.7386 | 0.79818 | 0.79426 | 0.79039 |
| 0.8114 | 0.80650 | 0.80263 | 0.79879 |
| 0.8667 | 0.81254 | 0.80868 | 0.80486 |
| 0.9120 | 0.81734 | 0.81350 | 0.80969 |
| 2-undecanone (1) + octane (2) | | | |
| 0.0452 | 0.70978 | 0.70579 | 0.70172 |
| 0.0929 | 0.71687 | 0.71289 | 0.70884 |
| 0.1347 | 0.72298 | 0.71901 | 0.71497 |
| 0.1971 | 0.73191 | 0.72796 | 0.72394 |
| 0.2448 | 0.73859 | 0.73465 | 0.73065 |
| 0.2953 | 0.74551 | 0.74158 | 0.73759 |
| 0.3464 | 0.75234 | 0.74842 | 0.74444 |
| 0.4014 | 0.75947 | 0.75557 | 0.75161 |
| 0.4521 | 0.76588 | 0.76199 | 0.75805 |
| 0.4993 | 0.77172 | 0.76785 | 0.76392 |
| 0.5542 | 0.77833 | 0.77447 | 0.77055 |
| 0.6021 | 0.78399 | 0.78013 | 0.77622 |
| 0.6534 | 0.78991 | 0.78606 | 0.78216 |
| 0.7133 | 0.79661 | 0.79278 | 0.7889 |
| 0.7550 | 0.80118 | 0.79735 | 0.79348 |
| 0.8077 | 0.80681 | 0.80298 | 0.79912 |
| 0.8558 | 0.81183 | 0.80802 | 0.80416 |
| 0.9052 | 0.81686 | 0.81306 | 0.80922 |
| 2-undecanone (1) + decane (2) | | | |
| 0.0359 | 0.73349 | 0.72970 | 0.72592 |
| 0.0941 | 0.73910 | 0.73531 | 0.73152 |
| 0.2064 | 0.75007 | 0.74626 | 0.74246 |
| 0.2482 | 0.75414 | 0.75033 | 0.74653 |
| 0.3308 | 0.76220 | 0.75838 | 0.75459 |
| 0.4053 | 0.76947 | 0.76565 | 0.76186 |
| 0.4301 | 0.77188 | 0.76806 | 0.76427 |
| 0.4542 | 0.77422 | 0.77039 | 0.76661 |
| 0.4913 | 0.77781 | 0.77399 | 0.77019 |
| 0.5383 | 0.78235 | 0.77853 | 0.77473 |
| 0.5867 | 0.78700 | 0.78318 | 0.77939 |
| 0.6934 | 0.79723 | 0.79341 | 0.78961 |
| 0.8155 | 0.80882 | 0.80500 | 0.80120 |
| 0.8664 | 0.81361 | 0.80979 | 0.80600 |
| 0.8972 | 0.81650 | 0.81268 | 0.80889 |
| 0.9703 | 0.82330 | 0.81948 | 0.81570 |
| 2-undecanone (1) + dodecane (2) | | | |
| 0.0558 | 0.75257 | 0.74891 | 0.74528 |
| 0.1032 | 0.75572 | 0.75205 | 0.74840 |
| 0.1540 | 0.75917 | 0.75550 | 0.75184 |
| 0.2022 | 0.76252 | 0.75884 | 0.75518 |
| 0.2511 | 0.76599 | 0.76230 | 0.75863 |
| 0.3050 | 0.76987 | 0.76617 | 0.76248 |
| 0.3524 | 0.77334 | 0.76962 | 0.76593 |
| 0.4175 | 0.77822 | 0.77449 | 0.77079 |
| 0.4524 | 0.78088 | 0.77714 | 0.77343 |
| 0.5004 | 0.78458 | 0.78082 | 0.77710 |
| 0.5682 | 0.78990 | 0.78614 | 0.78240 |
| 0.6464 | 0.79615 | 0.79237 | 0.78863 |
| 0.7111 | 0.80140 | 0.79762 | 0.79387 |
| 0.7845 | 0.80749 | 0.80370 | 0.79994 |
| 0.8498 | 0.81304 | 0.80924 | 0.80547 |
| 0.8990 | 0.81728 | 0.81347 | 0.80969 |
| 0.9498 | 0.82172 | 0.81790 | 0.81411 |
| 2-undecanone (1) + tetradecane (2) | | | |
| 0.0611 | 0.76557 | 0.76199 | 0.75848 |
| 0.0991 | 0.76735 | 0.76379 | 0.76026 |
| 0.1503 | 0.76988 | 0.76629 | 0.76275 |
| 0.2028 | 0.77253 | 0.76893 | 0.76539 |
| 0.2511 | 0.77507 | 0.77145 | 0.76789 |
| 0.2982 | 0.77764 | 0.77399 | 0.7704 |
| 0.3505 | 0.78056 | 0.77692 | 0.77333 |
| 0.3996 | 0.78342 | 0.77976 | 0.77615 |
| 0.4508 | 0.78648 | 0.78281 | 0.77918 |
| 0.5140 | 0.79039 | 0.78670 | 0.78306 |
| 0.5476 | 0.79253 | 0.78883 | 0.78518 |
| 0.5996 | 0.79595 | 0.79223 | 0.78857 |
| 0.6492 | 0.79931 | 0.79558 | 0.79191 |
| 0.7402 | 0.80571 | 0.80196 | 0.79825 |
| 0.8465 | 0.81368 | 0.80989 | 0.80615 |
| 0.8992 | 0.81781 | 0.81401 | 0.81028 |
| 0.9507 | 0.82200 | 0.81818 | 0.81443 |

[a] The standard uncertainties are: $u(T)$ = 0.01 K; $u(p)$ = 10 kPa; $u(x_1)$ = 0.0005. The relative combined expanded uncertainty for density (0.95 level of confidence) is $U_{rc}(\rho) = 0.002$.

**Table S4**

Excess molar functions: enthalpies ( $H_{\mathrm{m}}^{\mathrm{E}}$ ), volumes ( $V_{\mathrm{m}}^{\mathrm{E}}$ ) and isochoric internal energies ( $U_{V\mathrm{m}}^{\mathrm{E}}$ ) for 2-alkanone (1) + *n*-alkane mixtures at equimolar composition, 298.15 K and 95 kPa.

| *n*-alkane | $H_{\mathrm{m}}^{\mathrm{E}}$ /J mol$^{-1}$ | $V_{\mathrm{m}}^{\mathrm{E}}$ /cm$^3$ mol$^{-1}$ | $U_{V\mathrm{m}}^{\mathrm{E}}$ /J mol$^{-1}$ | $H_{\mathrm{m1}}^{\mathrm{E},\infty}$ /kJ mol$^{-1}$ | $U_{V\mathrm{m1}}^{\mathrm{E},\infty}$ /kJ mol$^{-1}$ |
|---|---|---|---|---|---|
| 2-butanone (1) + *n*-alkane (2) | | | | | |
| heptane | 1338 [S1] | 0.803 [S2] | 1113 | 7.5 [S1] | 6.1 |
| octane | 1408 [S1] | 0.887 [S3] | 1152 | | |
| decane | 1545 [S1] | 0.952 [S2] | 1261 | | |
| dodecane | 1658[S4] | 0.996 [S2] | 1356 | | |
| 2-hexanone (1) + *n*-alkane (2) | | | | | |
| hexane | 949 [S5] | 0.154 [S6] | 908 | | |
| heptane | 1055 [S5] | 0.375 [S6] | 947 | 6.6 [S5] | 5.7 |
| octane | 1132 [S5] | 0.516 [S6] | 981 | | |
| decane | 1268 [S5] | 0.643 [S6] | 1074 | | |
| 2-undecanone (1) + *n*-alkane (2) | | | | | |
| heptane | 587 | −0.192 | 646 | 4.1 | 4.2 |
| octane | 638 | −0.016 | 643 | | |
| decane | 707 | 0.197 | 647 | | |
| dodecane | 770 | 0.310 | 674 | | |
| tetradecane | 849 | 0.392 | 725 | | |

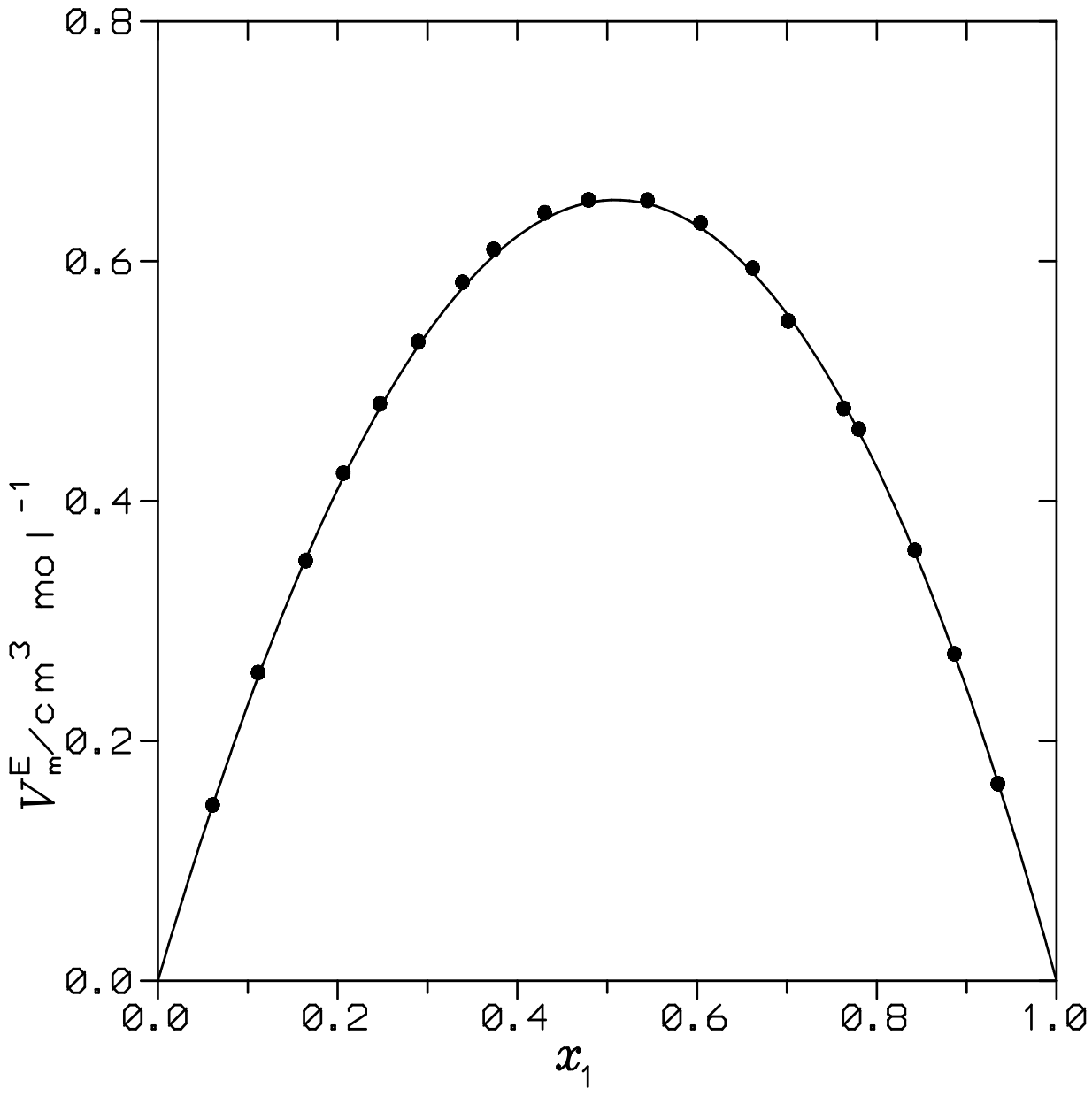

**Figure S1** $V_{\mathrm{m}}^{\mathrm{E}}$ of the cyclohexane (1) + benzene (2) mixture at 298.15 K and 95 kPa. Points, experimental results (this work). Solid lines, results from [S7].

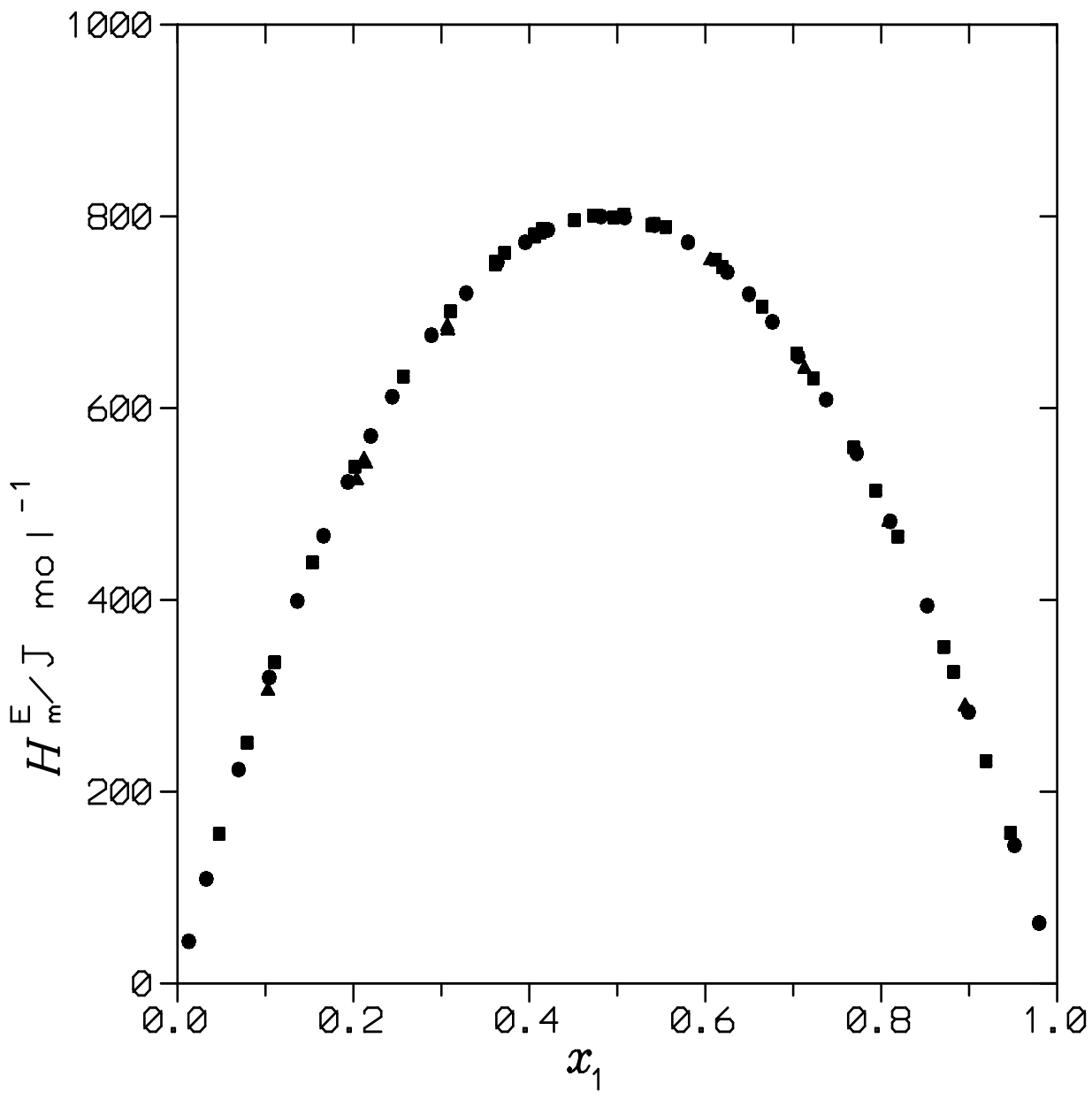

**Figure S2** $H^{\mathrm{E}}_{\mathrm{m}}$ of the cyclohexane (1) + benzene (2) mixture at 298.15 K and 95 kPa. Points: (▲), this work; (●), [S8]; (■), [S9].

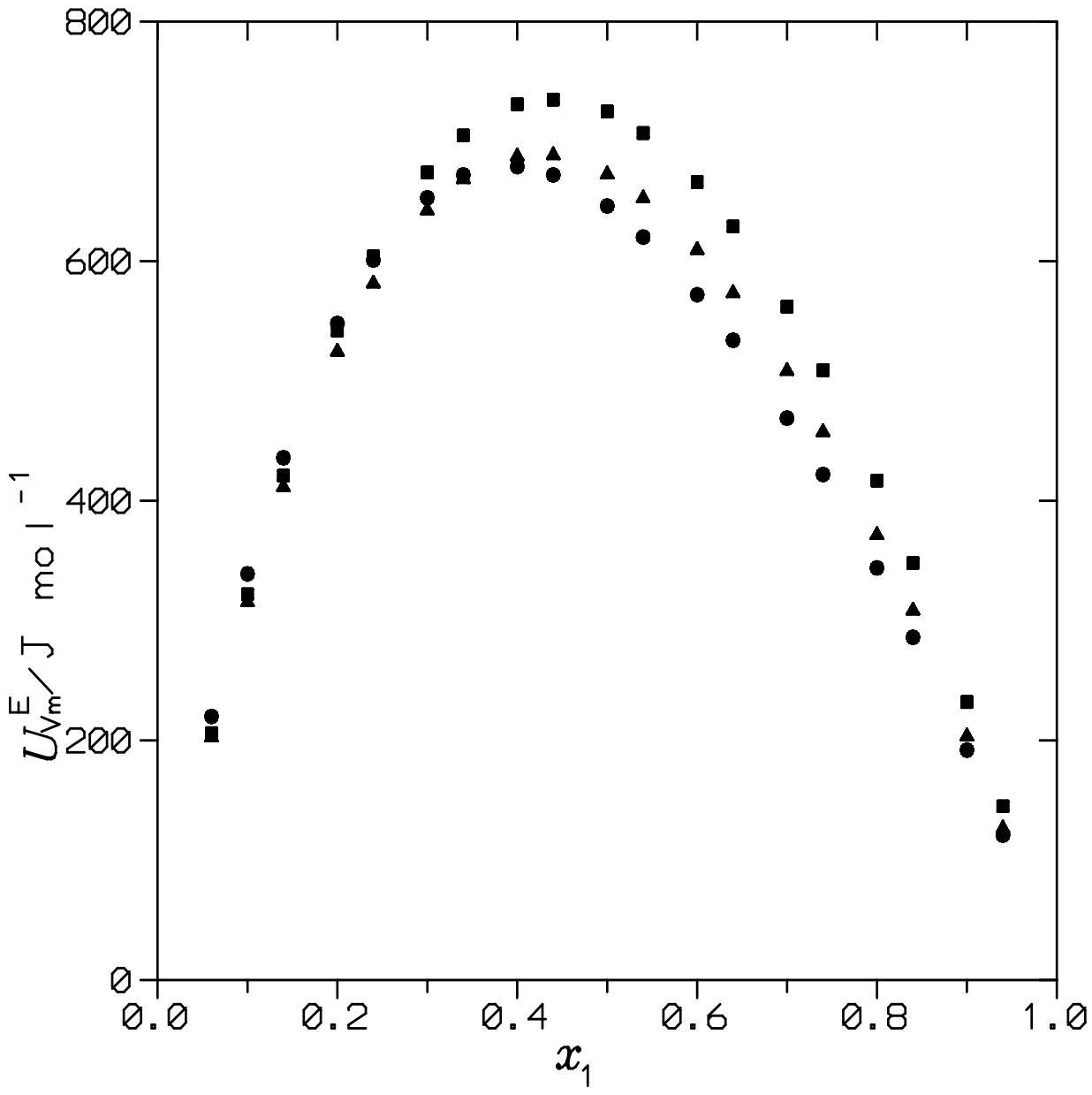

**Figure S3** $U^{\mathrm{E}}_{V\mathrm{m}}$ of 2-undecanone (1) + *n*-alkane (2) mixtures at 298.15 K and 95 kPa. Points, experimental results (this work): (●), heptane; (▲), dodecane; (■), tetradecane.